# Improving Services Offered by Internet Providers by Analyzing Online Reviews using Text Analytics

Suchithra Rajendran[a,b,*] and John Fennewald[a,b]

[a]Department of Industrial and Manufacturing Systems Engineering, University of Missouri Columbia, MO 65211, USA
[b]Department of Marketing, University of Missouri Columbia, MO 65211, USA

[*]Corresponding Author:
Suchithra Rajendran
E-mail address: RajendranS@missouri.edu
Telephone: 573-882-7421

# Improving Services Offered by Internet Providers by Analyzing Online Reviews using Text Analytics

## Abstract

**Purpose:** With the proliferation of digital infrastructure, there is a plethora of demand for internet services, which makes the wireless communications industry highly competitive. Thus internet service providers (ISPs) must ensure that their efforts are targeted towards attracting and retaining customers for continued growth.

**Design/methodology/approach:** Since Web 2.0 has gained traction and more tools have become available, customers in recent times are equipped to make well-informed decisions, specifically due to the colossal information available in online reviews. ISPs can use this information to better understand the views of the customers about their products and services. The goal of this paper is to identify the current strengths, weaknesses, opportunities, and threats (SWOT) of several leading ISPs by exploring consumer reviews using text analytics. The proposed approach consists of four different stages: bigram and trigram analyses, topic identification, SWOT analysis and Root Cause Analysis (RCA).

**Findings:** For each ISP, we first categorize online reviews into positive and negative based on customer ratings and then leverage text analytic tools to determine the most frequently used and co-occurring words in each categorization of reviews. Subsequently, looking at the positive and negative topics in each ISP, we conduct the SWOT analysis as well as the RCA to help companies identify the internal and external factors impacting customer satisfaction.

**Originality/value:** This research is one of the first to provide managerial insights for ISPs to improve the wireless services offered to their customers by analyzing online user reviews. We use a case study to illustrate the proposed approach. The managerial insights that are derived from the results can act as a decision support tool for ISPs to offer better products and services for their customers.

## 1. Introduction

In the United States, the wireless communications industry is proliferating with a reported growth of 4.1% from 2013 – 2018 and expected growth of 3.5% from 2018 to 2023 (Miller, 2018). This expected industry growth will lead to an estimated revenue increase of $114.6 billion (Hoffman, 2018; Miller, 2018). Several components factor into this anticipated advancement including faster connectivity speeds and better service, which leads to an overall desire for increased internet usage and the idea that internet access is a necessity rather than viewing it was a luxury (Miller, 2018). The wireless communications industry is highly competitive, as approximately two-thirds of the adults use broadband internet at home (Kohut, 2019).

Despite this level of competition, the industry is dominated by very few noticeable players, as the five biggest companies combine to control over 68% of total market share - AT&T (24.9%), Comcast (17.2%), Charter (12.9%), Verizon (7.4%), and CenturyLink (5.8%) (Miller, 2018). In such a competitive environment and a quickly growing industry, Internet Service Providers (ISPs) must ensure that their efforts are targeted towards attracting and retaining customers for continued growth of their business. Nearly 78% of revenue originates in the ISP-to-customer sector, representing the largest single market in which ISPs operate and compete. This sector includes both residential and business services with the most common activities involving downloading multimedia contents, browsing the internet, and using basic internet applications such as accessing

email accounts (Miller, 2018). ISPs in this domain compete on many factors, namely, availability of service, price, speed of connectivity and customer service (Hoffman, 2018).

In order to attract and retain the most customers possible, ISPs must determine the client expectations, and tailor their offerings to meet consumer needs and wants. One of the best ways for ISPs to identify what their users perceive about their services is to analyze online reviews posted by customers. Several social networking sites allow consumers to discuss their experiences with various ISPs and detail their levels of satisfaction or dissatisfaction. As Web 2.0 has gained traction and popularity, this style of content that is generated by users has increased rapidly (Guo et al., 2017). Though online customer reviews (OCR) suffers from the drawback that it is extremely subjective and only captures a small portion of the customer population, it is considered the second most reputable way that companies can extract user feedback (Nielsen Holdings, 2012). Reviews can be either negative or positive, and since negative reviews circulate five to six times faster than positive reviews, their impact on the business can be monumental (Salehan and Kim, 2016).

As social media usage continues to grow, enterprises experience additional roadblocks since they must examine OCR to understand how customers are interacting with their products and to determine the performance of their products and services (Sen and Lerman, 2007). Recently, studies have investigated online customer reviews (OCR), and shown the strengths and weaknesses associated with the products and services offered by a business (Salehan & Kim, 2016; Srinivas and Rajendran, 2019). By using OCR to understand the strengths and weaknesses of their offerings, any organization, including ISPs, can discover more ways to improve and enhance the experiences that customers have with their products and services.

Insights that ISPs may extract from online customer reviews include deciding which of their services to improve in order to satisfy their clients who have been disappointed in the past or determining the areas in which they perform very well and developing marketing plans to highlight those fields of service to attract more customers. As studies, such as Cheung et al. (2012), have shown that new customers place heavy emphasis on the opinions of existing clients and that OCR weigh heavily into a consumer’s view on which company to choose, it is extremely likely that potential ISP consumers might check OCR to gain insight into the experiences that prior users have had with each ISP that they are considering.

Each ISP adds value to customers in a different way tailored to its own business model. Many companies find success through economies of scale by bundling phone, internet, and television services into one package for consumers. Other ISPs find a niche market and focus on providing the best service in a particular sector, such as rural areas (Hoffman, 2018). Regardless of what an ISP’s business model is, OCR carry abundant information which can be beneficial to ISPs as they seek to provide the most value for their customers. The purpose of this paper is to identify the strengths, weaknesses, opportunities, and threats of an ISP utilizing knowledge mining to improve customer satisfaction based on information gained from online consumer reviews. Each consumer review explores various topics (e.g., speed, connection, customer service, etc.), and conveys an attitude (positive, negative, or neutral) towards each topic. In this study, we use text analytic methods to extract and analyze thousands of reviews for identifying those key topics and propose managerial recommendations based on our results.

The rest of the paper is organized as follows. Section 2 covers a review of several prominent studies discussed in the literature. A thorough discussion of the proposed methodological approach is given in Section 3. Section 4 includes the description of the case study as well as an analysis of the results obtained. Lastly, Section 5 presents the conclusions and the scope of later work.

## 2. Literature Review

As discussed earlier, the focus of this paper is to capture the voice of customers using online review mining and enable ISPs to provide better service. Therefore, in this section, we present some prominent studies related to wireless service providers and the impact of word of mouth on companies.

### 2.1 Wireless Service Providers

One of the very first studies on customer expectations in wireless service industry was pioneered by Zander (1997). The author explored the idea that wireless communications customers expect to receive the same wireless services as they would receive from a fixed network. The paper analyzed both a "universal" coverage scenario providing wideband services at all locations and a "hot-spot" scenario where coverage is offered, but only in select geographical areas. The study showed companies the benefits of presenting wireless services to all consumers and that as bandwidth increases, the infrastructure cost of a wireless system will also increase. Bar and Park (2005) looked at public Wi-Fi networks in the United States and the policy issues that the new trends raise. Metropolitan cities already have the existing infrastructure which supports the Wi-Fi networks, and they can help provide connectivity for city employees while making urban areas more desirable for residents and businesses.

With the rapid development in technology, broadband connectivity has become more widespread and voice services are more mobile, and as a result, ISPs are bundling voice services with broadband connectivity in homes. According to Markendahl and Makitalo (2007), ISPs consider the following important factors: network operation and access provisioning, customer acquisition and relations, and having a trust relationship with the customer. The authors also hypothesize that actors from outside the telecommunication industry will venture into this business and that wireless access is increasingly becoming a service with a wide coverage area rather than existing solely in the location where the service is officially provided.

Seo et al. (2008) highlighted the challenges that the wireless telecommunication market is facing, especially with the growing competition. Unlike goods, wireless services have to be offered continuously, and hence, developing strategies to retain and attract new customers is essential. Their study tried to understand how variables such as customer-related factors (e.g., age and gender) affect service plan complexity. A similar research was conducted by Quach et al. (2016), in which the authors examined the different aspects of an ISP's service quality and their impact on customer loyalty. Data was collected from more than 1200 internet users and segmented the information based on the customer's internet usage and recorded their perception of service quality dimensions. Their study proved that the service quality dimension has an impact on both attitudinal and behavioral loyalty.

### 2.2 Electronic Word of Mouth (eWOM)

Erkan and Evans (2016) examined the influence of eWOM in social media on consumers' purchasing intentions. Several hypotheses were considered and the authors proved that eWOM is positively related to consumers' purchasing intention, the usefulness of eWOM information is positively associated with the adoption of eWOM information, and quality of eWOM is positively related to usefulness. Wang et al. (2016) showed several differences between traditional WOM and eWOM, including the anonymity of eWOM messages, the fact that multiple people can access eWOM at any time, and that eWOM is more persistent and measurable. Their research indicated that about 70% of 28,000 internet users in 56 countries rely on online consumer reviews for recommendations.

Application of text analytic tools for examining eWOM has not only been used to study product improvement, but also in the service sector. For instance, Liang et al. (2018) studied the effect of eWOM on hospitality and tourism management. Their study showed that eWOM is highly influential due to its speed, amplitude and convenience. Likewise, Nielson (2007) claimed that most consumers perceive online opinions to be as trustworthy as brand websites. This idea, coupled with the fact that WOM reduces a company's ability to influence consumers through traditional marketing and advertising, shows how vital it is that companies become actively involved in online consumer communities.

While the studies mentioned above examined the relationship between online reviews and customer purchasing intentions, Javilanda et al. (2011) investigated the reason for the extensive spread of WOM. Their research showed that consumers spread WOM for several reasons including extreme satisfaction or dissatisfaction, commitment to the firm, length of the relationship with the firm and novelty of the product. Additionally, users may be motivated to share their experiences due to satisfaction, pleasure or sadness. The authors showed that the main antecedents of WOM influence are tie strength, demographic similarity, and perceptual affinity, and that WOM could influence product evaluations.

Balaji et al. (2016) examined how negative WOM (NWOM) on social networking sites (SNS) can adversely impact companies. Customer complaints can lead to brand dilution, volatile stock prices, and on a grand scale can even be the cause for public relations crises. Approximately 77% of online shoppers rely on user reviews to make purchasing decisions, and over one million people read product or service reviews every week on SNS, and more than 80% of these reviews are negative. Their study looked to the cognitive dissonance and social support theories to understand the determinants of NWOM on SNS. Since sharing opinions is a social activity, the social support theory shows that NWOM sharing may be due to the desire for social interaction. Additionally, consumers may engage in NWOM communications in order to reduce their cognitive dissonance levels because they are frustrated with a product or service.

This proposed research is one of the first to analyze online customer reviews posted for internet service providers using text analytics. We propose insights and managerial recommendations based on strategic planning tool (SWOT) and six sigma techniques (root cause analysis) to capture the voice of the customers. Though the literature discusses few studies that are conducting the SWOT analysis for internet service providers, to the best of the authors' knowledge, this study is the first to leverage customers' reviews and understand the voice of the customers.

**3. Methodology**

Figure 1 presents a flow diagram showing the methodological framework used for data collection and analysis. The proposed approach will first use a web scraper to extract thousands of publicly available customer review archives from online sources. Next, the reviews are separated into three categories based on the consumer's rating (which is on a scale of 1-5). On the rating scale, a value of "1" indicates a highly-dissatisfied customer, and "5" indicates a highly-satisfied customer.

All reviews with a rating of "1" or "2" are marked as "negative", reviews with a score of "3" are labeled as "neutral", and reviews with a rating of "4" or "5" are labeled as "positive". Upon separation into negative, neutral, and positive categories, bigrams and trigram analyses are used to identify the bag of words commonly co-occurring and gain further insights into the context in which consumers use certain terms. These frequently co-occurring words are analyzed and compared across different companies in the ISP industry with the goal of creating a customer-centric SWOT and root cause analysis using the negative, neutral, and positive bigrams and trigrams. Figure 1 provides an overview of the proposed approach.

PLEASE INSERT FIGURE 1 HERE

### 3.1 Web Scraping and Text Pre-Processing

The consumer reviews of the ISPs under study are scraped from several online customer review sites, and since the customer feedback might not be particularly centered around any one single aspect of the corporation, reviews are separated into individual sentences which are analyzed as independent comments. The reviews are cleaned prior to doing the text analysis by removing incomplete, duplicate and irrelevant reviews. Subsequently, the sentences are tokenized, special characters and non-English words are eliminated, inflected words are stemmed, characters are converted to lowercase, and stop and infrequent words are removed.

### 3.2 Topic Identification using Bigrams and Trigrams Analyses

After scraping and pre-processing, the most frequent topics discussed in reviews need to be identified, which is achieved by analyzing the frequent words, bigrams and trigrams. A bigram is a combination of two words that are used consecutively in relevance to the same topic. Examples pertaining to our study would be “customer service” or “connectivity issues”. A trigram is a group of three words that are used consecutively, and examples would be “good customer service” and “connectivity issues rural”.

This section presents an analysis of bigrams and trigrams, as discussed by Jurafsky and Martin (2014). If $x_1, x_2, \dots, x_N$ are a set of words, then the probability that these words occur in a sequence is denoted by $\mathrm{P}(X)$ as shown in Equation (1).

$$\mathrm{P}(X) = \mathrm{P}(x_1 \dots x_N) \tag{1}$$

The probability of $x_2$ occurring after $x_1$ in a sequence is shown in Equations (2) and (3). Similarly, the probability of $x_3$ occurring after $x_2$ and $x_1$ sequentially is given in Equation (4), and Equation (5) is derived from Equation (4).

$$\mathrm{P}(x_2|x_1) = \frac{P(x_1, x_2)}{P(x_1)} \tag{2}$$

$$P(x_1, x_2) = \mathrm{P}(x_2|x_1) \times P(x_1) \tag{3}$$

$$P(x_1, x_2, x_3) = \mathrm{P}(x_3|x_1, x_2) \times P(x_1, x_2) \tag{4}$$

$$P(x_1, x_2, x_3) = P(x_1) \times \mathrm{P}(x_2|x_1) \times \mathrm{P}(x_3|x_1, x_2) \tag{5}$$

If we extrapolate this expression to $N$ words, the probability of word $x_N$ occurring in a word sequence $x_1, x_2, \dots, x_{N-1}$ is given by Equations (6) and (7).

$$\mathrm{P}(x_1 \dots x_N) = P(x_1) \times \mathrm{P}(x_2|x_1) \times \mathrm{P}(x_3|x_1, x_2) \dots \mathrm{P}(x_N|x_1, x_2 \dots x_{N-1}) \tag{6}$$

$$\mathrm{P}(x_1 \dots x_N) = \prod_{n=1}^{\mathrm{N}} \mathrm{Pr}(x_n|x_1 \dots x_{n-1}) \tag{7}$$

Applying the conditional probability chain rule to the sequences of words under consideration, Equations (8) and (9) give the probability of $N$ words occurring in a sequence.

$$\mathrm{P}(x_1 \dots x_N) = P(x_1) \times \mathrm{P}(x_2|x_1) \times \mathrm{P}(x_3|x_1^2) \dots \mathrm{P}(x_N|x_1^{N-1}) \tag{8}$$

$$\mathrm{P}(x_1 \dots x_N) = \prod_{i=1}^{N} P(x_N | x_1^{N-1}) \tag{9}$$

In Equation (10), we consider the Markovian assumption for the bigram model.

$$P(x_N | x_1^{N-1}) = P(x_N | x_{N-1}) \tag{10}$$

where $P(x_N | x_{N-1})$ is approximated by using the proportion of the bigram count of $x_{N-1}$ and $x_N$ (represented by $\nu(x_{N-1}x_N)$) to the sum of the frequency of all the bigrams containing $x_{N-1}$ ($\sum_x \nu(x_{N-1}x)$) as shown by Equation (11).

$$P(x_N | x_{N-1}) = \frac{\nu(x_{N-1}x_N)}{\sum_x \nu(x_{N-1}x)} \tag{11}$$

Constraint (12) uses the trigram model to predict the conditional probability of the $N^{\text{th}}$ word in a sequence.

$$P(x_N | x_{N-1}, x_{N-2}) = \frac{\nu(x_{N-2}x_{N-1}x_N)}{\sum_x \nu(x_{N-2}x_{N-1}x)} \tag{12}$$

### 3.3 SWOT Strategic Planning using Bigrams and Trigrams Analyses

For each ISP, the reviews are separated based on their ratings as negative (<=2 stars), neutral (3-stars), and positive (>=4 stars), following which commonly co-occurring bigrams and trigrams are identified for each category. These bigrams and trigrams are then used to conduct a SWOT analysis for individual ISPs and the industry as a whole.

SWOT is a planning tool most commonly used for recognizing the strategic factors (both internal and external) that are important to the development of an enterprise (Kurttila et al., 2000). The strengths and weaknesses are identified by analyzing the internal characteristics of the organization, while opportunities and threats constitute the external elements, such as competition. The basic framework of SWOT analysis is shown in Figure 2. Such a SWOT assessment highlights the critical factors that influence the consumer's viewpoint of the corporation and can aid the ISP in their planning decisions.

PLEASE INSERT FIGURE 2 HERE

## 4. Case Study

We have chosen eight companies for this study that control over 70% of the total market share in the ISP domain. We have de-identified and referred to the ISPs under consideration as ISP#1 through #8. In total, 23,145 online customer reviews are considered with the following breakdown of reviews for each company: ISP#1 - 2407, ISP#2 - 4208, ISP#3 - 937, ISP#4 - 3446, ISP#5 - 989, ISP#6 - 3324, ISP#7 - 5019 and ISP#8 - 2815.

Table 1 shows a summary shows an overview of the ISPs considered and the primary services offered. As it can be seen, wireless internet is a primary service for every company except ISP#3 and ISP#6, and ISP#2 and ISP#3 are the only two companies that offer entertainment media as their primary services.

PLEASE INSERT TABLE 1 HERE

### 4.1 Data Description

Table 2 shows the data features and an explanation of each field. The two fields that are considered in this study are "Rating" and "Review". Figure 3 shows the distribution of reviews for the different internet service providers under consideration. We can see that the majority of the time, all companies are being negatively perceived by customers. More specifically, over 80% of the total reviews that are posted for ISP#1, ISP#3, ISP#4 and ISP#8 are given less than 2-star rating. On the other hand, relatively more positive reviews are observed for ISP#5, whereas neutral reviews are comparatively more in ISP#6.

PLEASE INSERT TABLE 2 HERE

PLEASE INSERT FIGURE 3 HERE

### 4.2. Experimental Results

Once the reviews are extracted, the data is cleaned using the procedure discussed in Section 3.1. This data pre-processing is done with the modules in the Python natural language toolkit (NLTK). In addition to the default English stop words in the NLTK toolkit, we also created a custom set of stop words to filter unnecessary words that do not add value to our study.

Table 3 gives an overview of the topics that occur most frequently in the OCR, a description of the topic and the significant words associated with each topic. Clearly, ‘customer service’, ‘connectivity’ and ‘speed’ are the key topics that customer reviews are emphasized on. For the purpose of illustration, the most frequently occurring bigrams and trigrams related to the ‘service’ Topic is shown in Figure 4. Many of these word combinations revolve around customer service or whether the actual services (wireless internet, TV, phone) are good or bad.

PLEASE INSERT TABLE 3 HERE

PLEASE INSERT FIGURE 4 HERE

Table 4 shows the list of most positively- and negatively-discussed topics based on bigram and trigram analyses. Topics related to customer service, connectivity, compatibility with other devices, billing and payments are several of the most common.

PLEASE INSERT TABLE 4 HERE

### 4.3 SWOT Analysis

From the topics discussed in Table 4, SWOT analysis for each ISP can be developed. A sample SWOT Analysis for ISP#1 is given in Table 5. Following that, we also propose several managerial insights based on the observations made.

PLEASE INSERT TABLE 5 HERE

#### 4.3.1 Managerial Insights for ISP#1

Though this section presents the managerial insights derived from SWOT analysis pertaining to a specific company (ISP#1), a similar approach can be adopted for other providers as well.

- *Customized Bundle Plans:* Develop a bundling plan that can be tailored to individual consumer's preferences. In other words, the most effective package will be the one which allows consumers to pay for specific services which they value and nothing more (e.g., fast speeds to play games online but limited TV channels, wide range of TV channels without high-speed internet).
- *Employee Training:* Ensure that all customer service representatives are well trained on how to best serve their customer base. This may include (i) educating them on products, policies, and packages that the company offers, and (ii) clearly communicating with customers on billing dates and plans.
- *Marketing Company Strengths:* Market the company's strongest suits such as high speed internet and widespread coverage. Moreover, as one of the industry leaders, it is essential to advertise that ISP#1 possesses a large portion of the market share as well.

### 4.4 Link and Root Cause Analyses

Figure 5 presents a positive link analysis representing the overall topics that are commonly discussed in the positive reviews across all ISPs. A larger-sized node indicates that the topic is spoken more often by customers. Clearly, we can see that ‘Bundling’, ‘Connectivity’, and ‘Technical Service’ were three of the most repeated topics in the positive review domain. Under the ‘bundling’ topic, customers praise the triple and quadruple play packages as well as the family plans due to their cost savings. The topic ‘Connectivity’ is observed to be positively spoken by customers of certain companies and negatively for other ISPs. Thus we interpret that connectivity is a key issue that companies must pay attention to. Customers prioritize consistent connection over speed, while good online gaming streaming is also a factor that they consider. During a technical issue, customers expect proper assistance either by visiting them at home or solving the problem over the phone (i.e., easy troubleshooting).

Figure 6 depicts a root cause analysis fishbone diagram highlighting the overall causes that lead to customer dissatisfaction in the internet service sector. The primary causes are ‘Contracts’, ‘Connectivity’, ‘Cell Phone Service’, and ‘Call Center’. The associated secondary causes are also shown in Figure 6. Clearly, customers are highly unsatisfied when they are charged with additional fees without prior notification. Interrupted service and slow loading speed in rural areas are some issues that users experience. Moreover, they are not comfortable with the automation, especially studies have proven that the older population are resistant to automation changes (Vaportzis et al., 2017).

PLEASE INSERT FIGURES 5 AND 6 HERE

### 4.5 Managerial Implications

Based on the bigram and trigram analyses, we suggest the following recommendations to the internet service providers as a whole.

- *Streamline customer service call process:* When customers attempt to contact ISPs over the phone, they often complain that it is difficult to speak to a human and that processes are automated. Our recommendation would be for ISPs to have online live chat programs that can achieve a balance between customer satisfaction and cost.
- *Provide positive work environment for customer service representatives:* Many online reviews mention frustrations with customer service representatives, specifically, them being "rude", "unhelpful" or "not caring" about the user problems. Prior studies have proven a positive relationship between employee job stress, job satisfaction, and customer loyalty (Loveman, 1998; Hansemark and Albinsson, 2004; Hill and Alexander, 2017). Therefore, we theorize that the rude customer services provided are due to the burnout that employees experienced from answering customer calls monotonously. Hence it is recommended that ISPs provide a more positive work environment offering free/discounted health club memberships and conduct regular medical checkups. Also, ISPs can develop a handbook for call center representatives providing a set on instructions on how to handle different types of customer grievances.
- *Payment communication:* Many consumers are upset that ISPs did not communicate clearly about contracts, charging extra fees, change in billing dates, etc. Our recommendation is for ISPs to interact and update customers by different modes of communication (such as phone calls, emails and text messages).

- *Rural area connectivity:* A large portion of customers complain about poor internet service in rural areas. We suggest that stronger infrastructure has to be established in the countryside to better serve their clients and provide faster speeds.
- *Device compatibility:* Some consumers criticize that their internet service does not work with various devices, whether that be their brand of routers, cellphones, or their gaming console. To avoid this issue, we recommend that ISPs keep device compatibility as a priority when designing their products and services.
- *Improve customer service representative language training:* Based on the reviews, consumers are commonly distressed that the representatives are unable to fluently speak in English and hence not being able to entirely comprehend their grievances. We recommend that ISPs offer extensive language training for their representatives who are non-native English speakers.

## 5. Conclusions

Studies have shown that online consumer reviews factor heavily into consumer's decisions across many industries, including ISPs. This research is one of the first to provide managerial insights for ISPs to improve the wireless services offered to their customers by analyzing online user reviews. While very few research has been conducted involving SWOT analysis from an internet provider's perspective, examining online customer reviews enables ISPs to understand their consumer needs better. Over 23,000 online reviews were first collected using web scraping, reviews were subsequently separated based on their star ratings following which commonly co-occurring words were identified in each category, and a SWOT and root cause analyses were finally conducted.

We identified six meaningful topics using our approach and conducted SWOT analysis to showcase the critical factors that impact consumers' choice of ISP. We also proposed the following managerial recommendations to the ISP sector, based on the results obtained.

- Streamline customer service call process
- Provide positive work environment for customer service representatives
- Interact with customers using different means of communication
- Establish proper infrastructure in rural areas to improve connectivity
- Make design changes to ensure device compatibility
- Provide language training for representatives

## LIST OF FIGURES

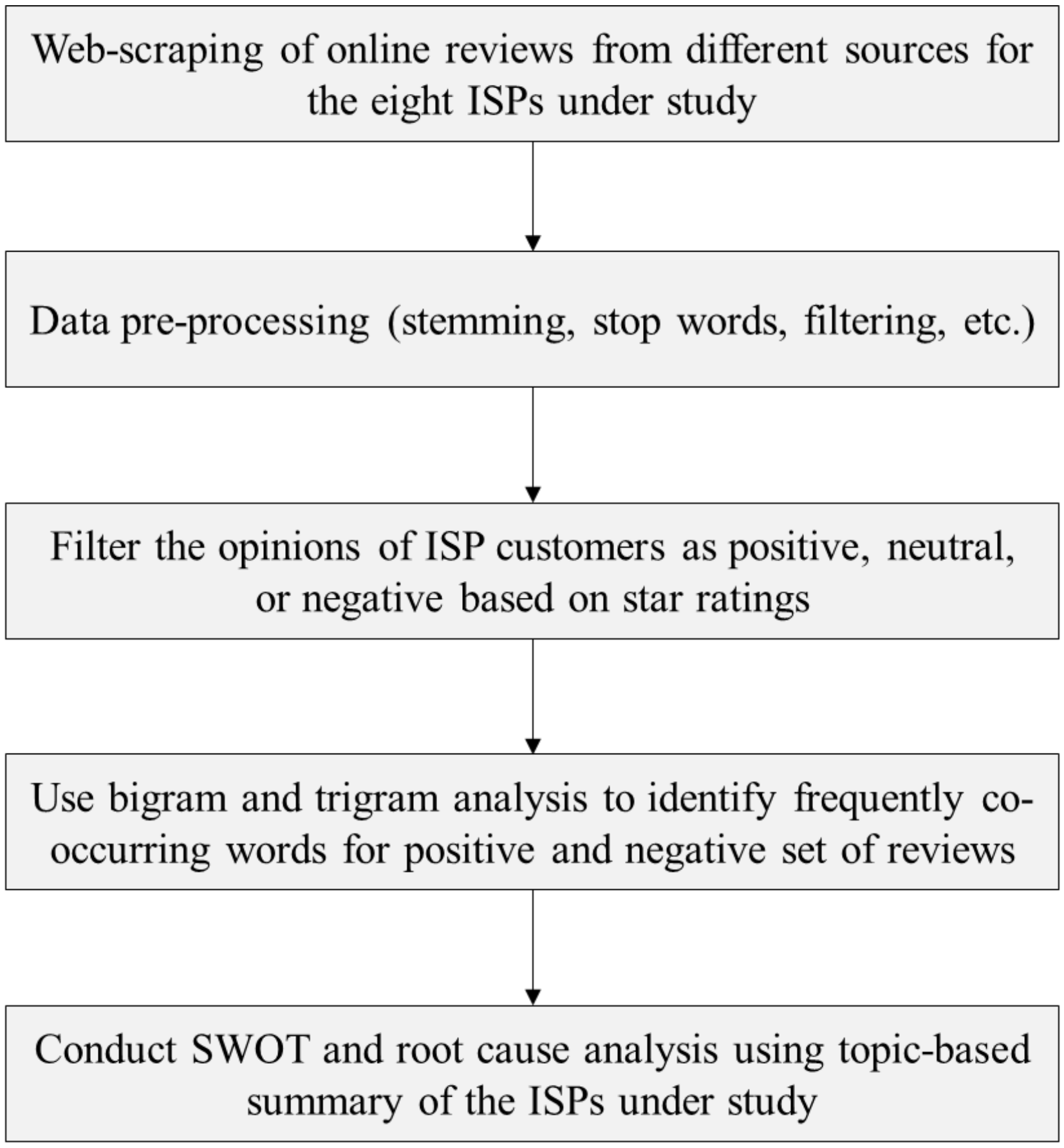

**Figure 1: Methodological Framework**

| | | |
|---|---|---|
| **Internal** | **Strengths**<br>ISP offerings that enrich customer's experience and lead to positive WOM (e.g., fast speeds, helpful technicians) | **Weaknesses**<br>Qualities of the ISP that hurt customer experience and create a negative perception (e.g., poor customer service, slow speeds in rural areas) |
| **External** | **Opportunities**<br>Potential changes companies can adapt to enhance reputation (e.g., bundling packages, better trained customer service representatives) | **Threats**<br>Obstacles that the ISP faces due to external environment (e.g., governmental regulation, competition from other companies) |

**Figure 2: Basic Framework of SWOT Analysis based on ISP Customer Experience**

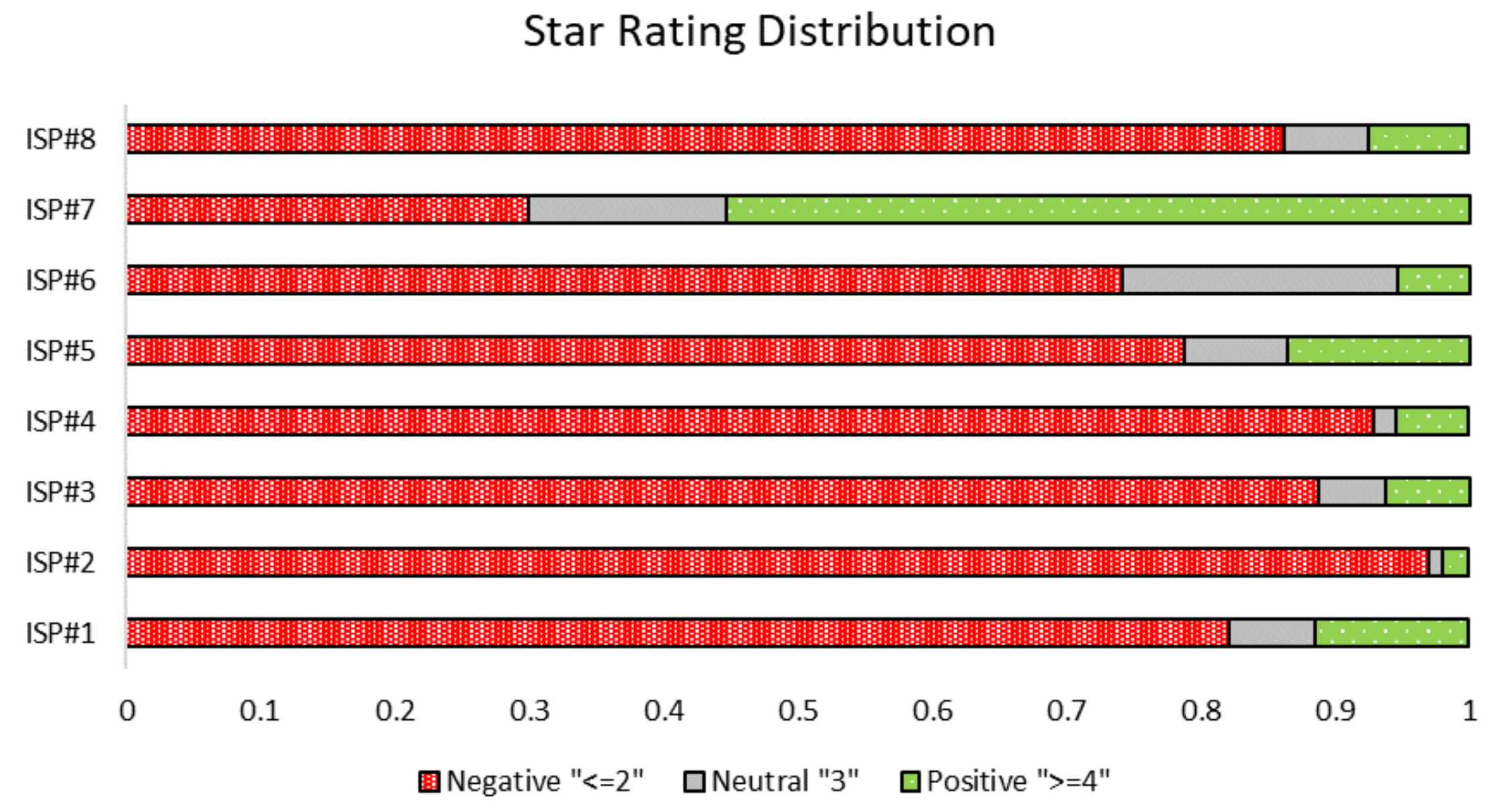

**Figure 3: Distribution of the Reviews**

BIGRAMS

(‘customer’, ‘service’), (‘cancel’, ‘service’), (‘phone’, ‘service’), (‘service’, ‘call’), (‘service’, ‘slow’), (‘tv’, ‘service’)

TRIGRAMS

(‘worst’, ‘customer’, ‘service’), (‘terrible’, ‘customer’, ‘service’), (‘contact’, ‘customer’, ‘service’), (‘poor’, ‘customer’, ‘service’), (‘horrible’, customer’, ‘service’), (‘call’, ‘customer’, ‘service’), (‘customer’, ‘service’, ‘rep’), (‘call’, ‘cancel’, ‘service’)

**Figure 4: Frequently Occurring Bi-grams and Tri-grams for Negative Opinions Related to the Topic ‘Service’**

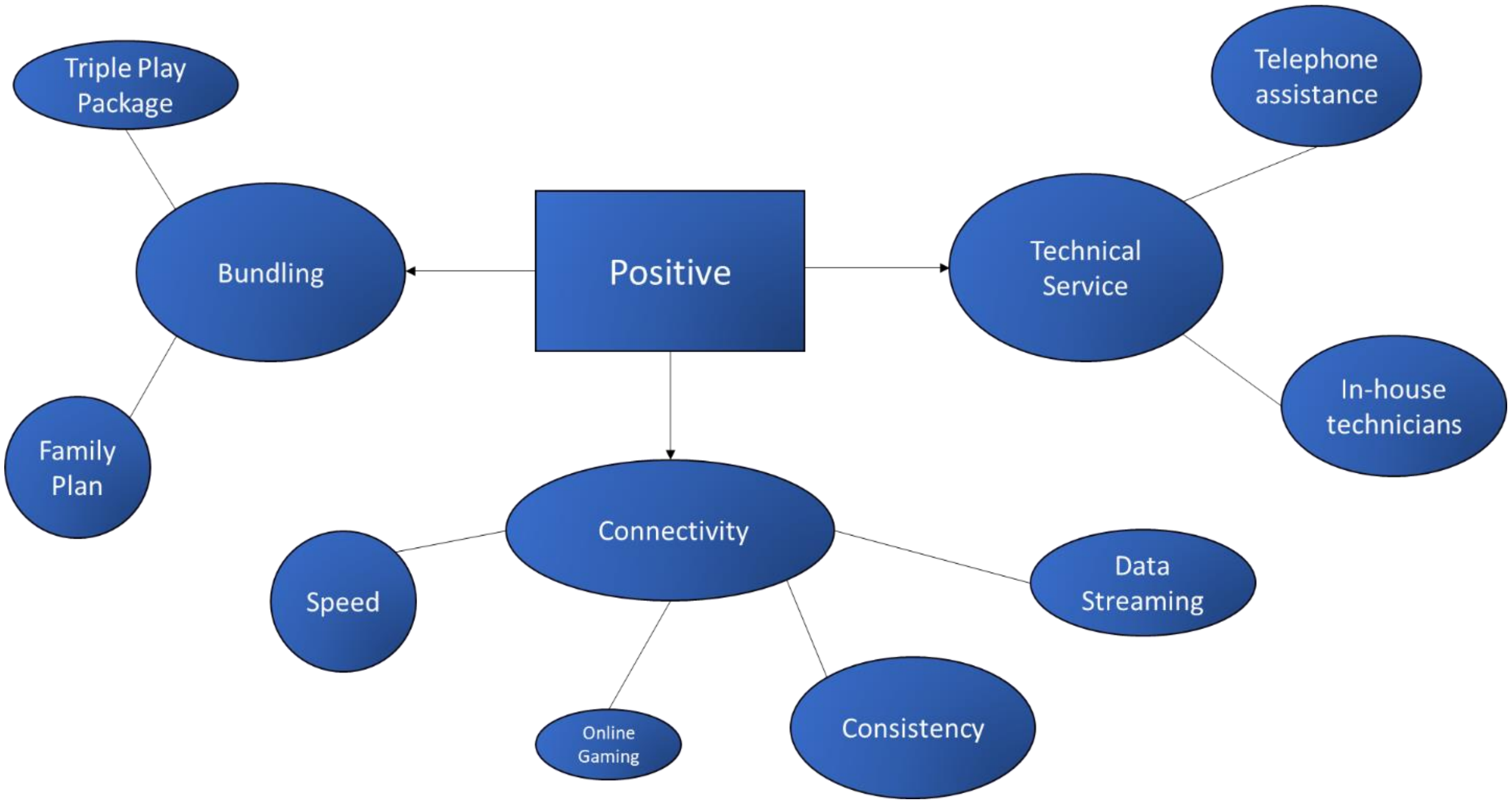

**Figure 5: Positive Link Analysis**

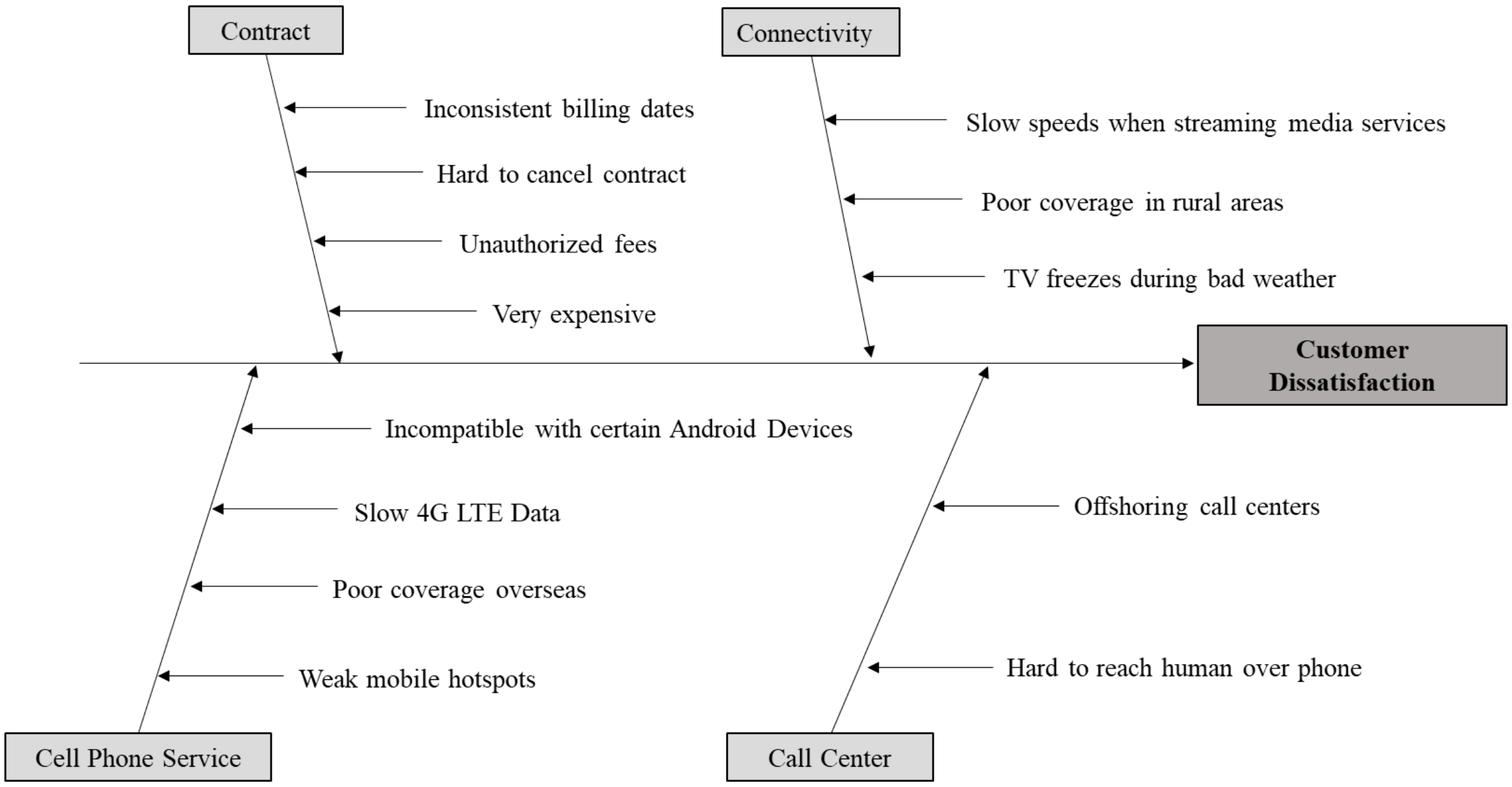

**Figure 6: Root Cause Analysis Ishikawa Diagram**

# LIST OF TABLES

**Table 1: Overview of the ISPs Considered**

| Company | Primary Services Offered | | |
|---|---|---|---|
| | Wireless Internet | Phone Services | Television |
| ISP#1 | x | x | |
| ISP#2 | x | | x |
| ISP#3 | | x | x |
| ISP#4 | x | | |
| ISP#5 | x | x | |
| ISP#6 | | x | |
| ISP#7 | x | x | |
| ISP#8 | x | | |

**Table 2: Data Fields**

| Field | Explanation |
|---|---|
| Rating | Contains a rating on a scale from 1-5 of how much the customer liked the service. The value “1” represents a highly unsatisfied customer while a “5” represents a highly satisfied customer |
| Review | Contains the review body |

**Table 3: Key Topics Overview**

| Topic | Description | Significant Words |
|---|---|---|
| Service | The service provided to customers by their ISP | Slow Installation, Setup Time |
| Phone | Using a phone with service from an ISP | Coverage, Connectivity, Cell, Data |
| Contract | The signed agreement between ISPs and their customers | Cancel, Terminate, Switch, Billing |
| Rural Area | ISP performance in sparsely populated areas | Rural Connection, Dead Zone |
| Representative | Employee of an ISP who helps customers with difficulties | House Calls, Technician, Unhelpful |
| Speed | Measure of how quickly service reaches customers | Fast, Slow, Download, High |

**Table 4: Positively and Negatively Perceived Topics by Customers for each ISP**

| ISP | Topics Based on Positive Reviews | Topics Based on Negative Reviews |
|---|---|---|
| ISP#1 | Fast speeds internet in cities; Versatile; Easy installation | Slow service in rural areas; Changing billing dates and switching plans without prior notification; Routers are not durable; Slow speed while streaming media |
| ISP#2 | Helpful technician; Good streaming for playing online games | Wrongful billing; Increasing service costs |
| ISP#3 | Triple Play Package; Fast speed; Wide TV channel variety | Hidden fees; Disrupted TV streaming |
| ISP#4 | Works well even during bad weather; Good upload and download speeds | Slow installation; Overseas call centers; Fixed 2-year contract; Poor transition from one generation to its next |
| ISP#5 | Good family plan; Cost effective; Service representative speaks in English fluently | Unauthorized payments and charges; Hard to cancel contract |
| ISP#6 | Binge data streaming; Good service in Europe; Service representatives give good instructions | Offshore call centers; Incompatible with certain Android devices |
| ISP#7 | Consistent internet service; Helpful salespeople | Poor international service; Call center representatives do not speak English fluently; Poor signal in rural areas; Slow 4G LTE |
| ISP#8 | Helpful technicians; Wide TV channel selection; Easy and quick troubleshoots; Providing house calls services | Services are priced high; DVR is expensive |

**Table 5: SWOT Analysis for ISP#1**

| Strengths | Weaknesses |
|---|---|
| - Fast speeds<br>- Quick installation<br>- Widespread coverage<br>- Highly compatible across different devices | - Routers are not durable<br>- Irregular payments<br>- Expensive<br>- Switching plans and billing dates<br>- Slow streaming speed of media-services |
| **Opportunities** | **Threats** |
| - Expand TV channel offering<br>- More comprehensive training for customer service representatives<br>- Improve services in rural areas | - Well marketed Triple and Quadruple Play packages by other ISPs<br>- House calls services offered by certain companies |